\documentclass[doublecol,figures]{epl2}

\usepackage{epsfig}

\title{Pattern Formation of Glioma Cells: Effects of Adhesion}

\author{Evgeniy Khain\inst{1} \and Casey M. Schneider-Mizell \inst{3} \and Michal O. Nowicki\inst{2} \and E. Antonio Chiocca\inst{2} \and S. E. Lawler\inst{2} \and Leonard M. Sander\inst{3}}

\institute{
\inst{1} Department of Physics, Oakland University, Rochester, Michigan 48309 \\
\inst{2} Department of Neurological Surgery, The Ohio State University Medical Center, Columbus, Ohio 43210 \\
\inst{3}Department of Physics and Michigan Center for Theoretical Physics, The University of Michigan, Ann Arbor, Michigan 48109
}


\abstract {We investigate clustering of malignant glioma cells. \emph{In vitro} experiments in collagen gels identified a cell line that formed clusters in a region of low cell density, whereas a very similar cell line (which lacks an important mutation) did not cluster significantly. We hypothesize that the mutation affects the strength of cell-cell adhesion. We investigate this effect in a new experiment, which follows the clustering dynamics of glioma cells on a surface. We interpret our results in terms of a stochastic model and identify two mechanisms of clustering. First, there is a critical value of the strength of adhesion; above the threshold, large clusters grow from a homogeneous suspension of cells; below it, the system remains homogeneous, similarly to the ordinary phase separation. Second, when cells form a cluster, we have evidence that they increase their proliferation rate. We have successfully reproduced the experimental findings and found that both mechanisms are crucial for cluster formation and growth.}

\begin{document}

\maketitle

The process of tumor growth has attracted a good deal of attention in the physics community in recent years \cite{Tracqui}. In addition to its medical importance, it presents an exciting example of pattern formation \cite{Deutsch} and collective cell behavior \cite{Couzin} in intrinsically non-equilibrium systems. One of the rapidly developing areas of tumor growth studies is a theoretical and computational investigation of glioblastoma multiforme (GBM) \cite{Nature}, the most malignant of primary brain tumors. Malignant gliomas are not treated effectively by current therapies \cite{CBTRUS03}. This is partly due to the fact that GBM is highly invasive \cite{Demuth2}. Indeed, glioma cells not only proliferate, but detach from the primary tumor and migrate away into the extracellular matrix. To mimic the \emph{in vivo} avascular tumor growth, the \emph{in vitro} growth of multicellular tumor spheroids is frequently considered \cite{spheroids}. In recent experiments, {\it in vitro} glioma growth (starting from a small tumor spheroid) was investigated in a transparent gel for two cell lines, U87WT (wild type) and U87$\Delta$EGFR \cite{Stein}. In the second case, there is a mutation that is known to lead to enhanced malignancy \cite{EGFR}. In the experiment, the growing spheroid of glioma cells (the proliferative zone) is surrounded by zone of low-density motile cells known as the invasive zone. This morphology is similar to what happens in the brain where deadly recurrent tumors form via motile invasive cells. In the experiments $\Delta$EGFR forms small clusters of cells in the invasive region, but WT does not \cite{Stein}. It is probable that the mutation in the $\Delta$EGFR line, which is known to make the EGF receptors constitutively active, also causes these cells to be more adhesive than the WT line, and this results in enhanced clustering. These clusters may give a clue to the formation of recurrent brain tumors. The investigation of the mechanism of the  clustering glioma cells is the subject of the present work.

In a growing tumor, the maximal proliferation activity occurs at the tumor border \cite{Bru,Giese}. Cells in the invasive and proliferative zones differ in phenotype: compared to cells in the inner proliferative zone, individual invasive cells have a much lower proliferation rate \cite{Giese}. This dichotomy between invasive cells (which rapidly migrate but rarely proliferate) and proliferative cells located on the tumor surface has been observed experimentally \cite{Giese} and addressed theoretically \cite{Khain0,Stein,Iomin}. We will incorporate this experimental observation in our modeling.

To investigate clustering, we did an experiment to follow the clustering of glioma cells on a surface. On a substrate, glioma cells migrate, proliferate and experience cell-cell adhesion. One can define a characteristic time for migration, $\tau$, as a time required for a cell to move its own diameter (of the order of $10 \mu m$). The motion of cells is an active (and highly complicated) process. Many experiments have shown that the motion is not purely diffusive (even in the absence of chemotaxis) but shows persistence on small scales \cite{persistent,Hegedus}. However, on length scales larger than the persistence length, random motion is a reasonable approximation. The characteristic migration time $\tau$ can be found from the analysis (tracking) of the cell trajectories on a substrate \cite{Hegedus}. Figure $1b$ in Ref. \cite{Hegedus} shows the displacement of glioma cells migration versus time. On scales much larger than the cell diameter, this motion looks like diffusion with an effective diffusion time of $4$ minutes. The effective migration time in our experiments is of the same order of magnitude. The typical time for proliferation is much larger, of the order of one day. However, the rate of proliferation depends on cell phenotype.

\begin{figure} [ht]
\vspace{-0.2cm} \centerline{
\begin{tabular}{cc}
\includegraphics[width=6cm,clip=]{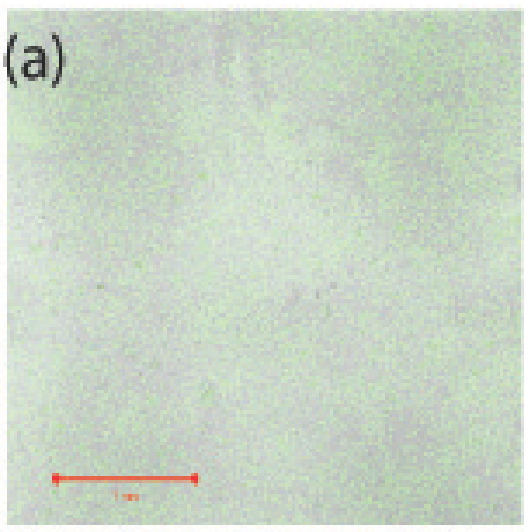}
\\
\includegraphics[width=6cm,clip=]{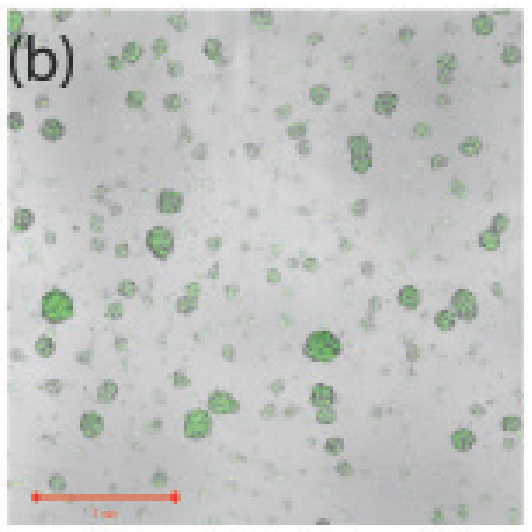}
\end{tabular}}
\caption{Snapshots of the system for the two cell lines five days after the beginning of the experiment \cite{exp}. $\Delta$EGFR cells
form clusters (b), while WT cells are homogeneously distributed over the system (a). The typical cell diameter is $10$ $\mu$m, so
each cluster in Fig.~\ref{snapshot}b contains hundreds of cells.}
\label{snapshot}
\end{figure}

\begin{figure}[ht]
\onefigure[width=6cm]{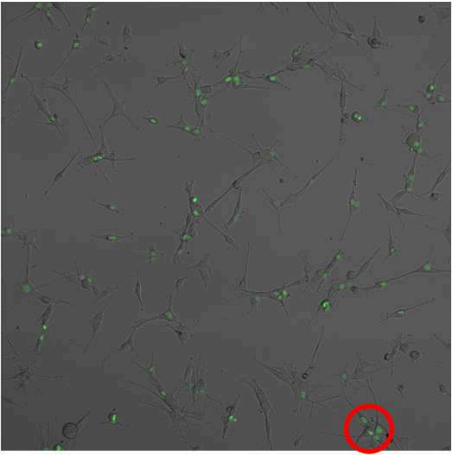}
\caption{$\Delta$EGFR cells a few hours after beginning of the experiment. Shown is a part of the system, $1400 \mu m$ on $1400 \mu m$. A circle shows a small cluster of cells, which randomly migrated and stuck together due to cell-cell adhesion.}
\label{snapshotnew}
\end{figure}

In the experiment, we directly compared a low-density preparation of U87WT and U87$\Delta$EGFR cell lines \cite{exp}. The cells were seeded on plastic Petri dishes at $50$ percent initial confluence. Images were taken every 24 hours with a Zeiss LSM510 confocal microscope. Bright field images were captured on a transmitted light detector; cells were marked with GFP (green). Figure \ref{snapshot} shows the experimental results $5$ days after the beginning of experiment. The $\Delta$EGFR cells form large clusters (b), while WT cells remain homogeneous (a). A typical cluster size in Fig.~\ref{snapshot}b is of the order of one hundred microns, so that each cluster contains hundreds of cells. Since cell-cell adhesion promotes clustering, the results suggest that $\Delta$EGFR cells are more adhesive than the wild type cells. We have measured the cluster size distribution for the $\Delta$EGFR; the results will be shown below. We examined the initial stages of the experiment in details. We made a movie, consisting of $50$ frames with $4$ minutes interval between the frames. We observed that very small clusters (that include just several cells) were formed when cells migrated, randomly approaching each other and stuck together, see Fig.~\ref{snapshotnew}.

In order to investigate these phenomena in detail we formulate a mathematical model for motile, proliferative and adhesive cells on a two-dimensional square lattice \cite{Khain1,Khain2}. Each lattice site can be empty or once occupied by a cell. We take the lattice distance to be equal to the cell diameter. The dynamics are as follows: a cell is picked at random. It can proliferate with probability $\alpha$. If it does not proliferate, we allow the possibility of migration to a neighboring site if the site is empty. The probability for migration decreases as the number of nearest neighbors increases. In our model it is given by $(1-\alpha)(1-q)^n$, where $0<q<1$ is the adhesion parameter, and $n$ is the number of nearest neighbors. This is a schematic representation of the underlying biology, but it is a convenient assumption. The case of no adhesion corresponds to $q=0$. For large $q$, adhesion makes it quite hard to move a cell that has many neighbors. As we mentioned above, the migration time $\tau$ (the time for a cell to move one cell diameter) is much shorter than the proliferation time (the time for one cell division). Therefore, $\alpha \ll 1$. After each cell process, the time is advanced by the migration time divided by the current number of cells. Numerical solutions of the model show that depending on the strength of cell-cell adhesion, we can get qualitatively different types of time evolution of the system. A similar model has been used in the context of wound-healing \cite{Khain1}.
\begin{figure}[ht]
\onefigure[width=7cm]{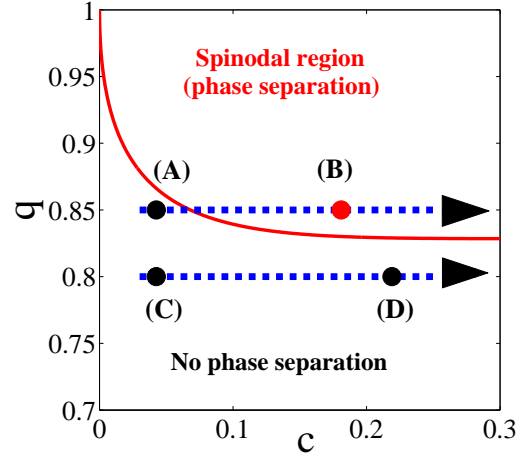}
\caption{A part of the phase diagram, for small densities and
large adhesion parameters. The probability of migration
decreases as the number of nearest neighbors increases to account
for cell-cell adhesion. A border of the phase separation region
(the solid line) is given by Eq.~(\ref{Onsager}). We start from
small density and turn on proliferation. The average density of
cells increases, and one can either enter the two-phase region
[path (A)-(B)] or remain in the stable region [path
(C)-(D)], depending on the adhesion parameter (two dotted lines).}
\label{phase_diagram}
\end{figure}

In the case of non-proliferating cells ($\alpha=0$), the existence of different regimes of clustering behavior is well known since
our scheme can be mapped into the Ising model \cite{Huang}: we identify  the adhesion parameter $q$ with
$1-\exp(-J/k_BT)$, where $T$ is the  temperature, $k_B$ is Boltzmann's constant, and $J$ is the coupling strength in the
magnetic model. The average density of cells, $c$, is identified with $(m+1)/2$ where $m$ is the magnetization. By exploiting this
mapping, we plot a phase diagram in terms of $q$ and $c$ for the case of no proliferation in Fig.~\ref{phase_diagram}. We present the part of the phase diagram corresponding to small $c$ and large $q$. In Fig.~\ref{phase_diagram}, the solid curve separates the phase diagram into two qualitatively different regions. Below the curve, an initially homogeneous ensemble of cells remains homogeneous. Due to non-zero adhesion some small-size clusters form, but these clusters do not grow. If the cell-cell adhesion parameter exceeds a critical value $q_{c}(c)$ determined by \cite{Huang}:
\begin{equation}
c = \frac{1}{2} \pm \frac{1}{2}\left[ 1 -
\frac{16(1-q_{c})^2}{q_{c}^4} \right]^{1/8}, \label{Onsager}
\end{equation}
we enter the region in the phase diagram \cite{Godreche} where the homogeneous state becomes unstable, and there is phase separation
and clustering.

For non-zero proliferation, the diagram shown in Fig.~\ref{phase_diagram}, gives only qualitative predictions, which we test numerically
(see below). Nevertheless, it allows us to propose two qualitatively different scenarios of time evolution and growth of cell population  depending on the strength of cell-cell adhesion. Consider an initially homogeneous system of cells with a very small density (points $A$ and $C$ in Fig.~\ref{phase_diagram}) and turn the proliferation on. The average density of cells increases (dotted lines), and the system either enters the two-phase region ($B$) or remains in the stable region ($D$), depending on $q$. For supercritical adhesion, a phase separation occurs between high density clusters and a ``gas" of single cells. Then these clusters start growing mostly by proliferation (since coarsening \cite{coarsening} is a very slow process). If $q$ is smaller than the threshold adhesion strength $q_{c}$, the system remains homogeneous, and the
growth of cell population is entirely determined by proliferation.

It turns out that the proliferation rate $\alpha$ is not constant: as we mentioned above, cells on the surface of tumor spheroids (these cells are called {\it proliferative}) divide much more often than single individual cells (which are called {\it invasive}). We incorporate this experimental observation in the model, assuming that cells located on the surfaces of clusters have an increased proliferation rate, which is denoted by $\alpha_{high}$. Notice that inside the cluster cells (which have no empty neighboring sites) can not proliferate.  Proliferation of invasive cells is assumed to depend on cell density: when the density is high, cells proliferate less due to contact inhibition effect. An experiment was done \cite{Chopp} to quantify the proliferation rate of invasive cells. We found that it can be fitted by $\alpha(n) = \alpha_{low}(1+n)^\beta (1-n)$ with $\beta=1.73$, where $0<n<1$ is the local area fraction and $\alpha_{low}$ is the basic proliferation rate for invasive cells. Invasive cells may switch their phenotype to proliferative, as they form sufficiently large cluster. Then their rate of division increases significantly from $\alpha_{low}$ to $\alpha_{high} \gg \alpha_{low}$.

\begin{figure} [ht]
\vspace{-0.2cm} \centerline{
\begin{tabular}{cc}
\includegraphics[width=6cm,clip=]{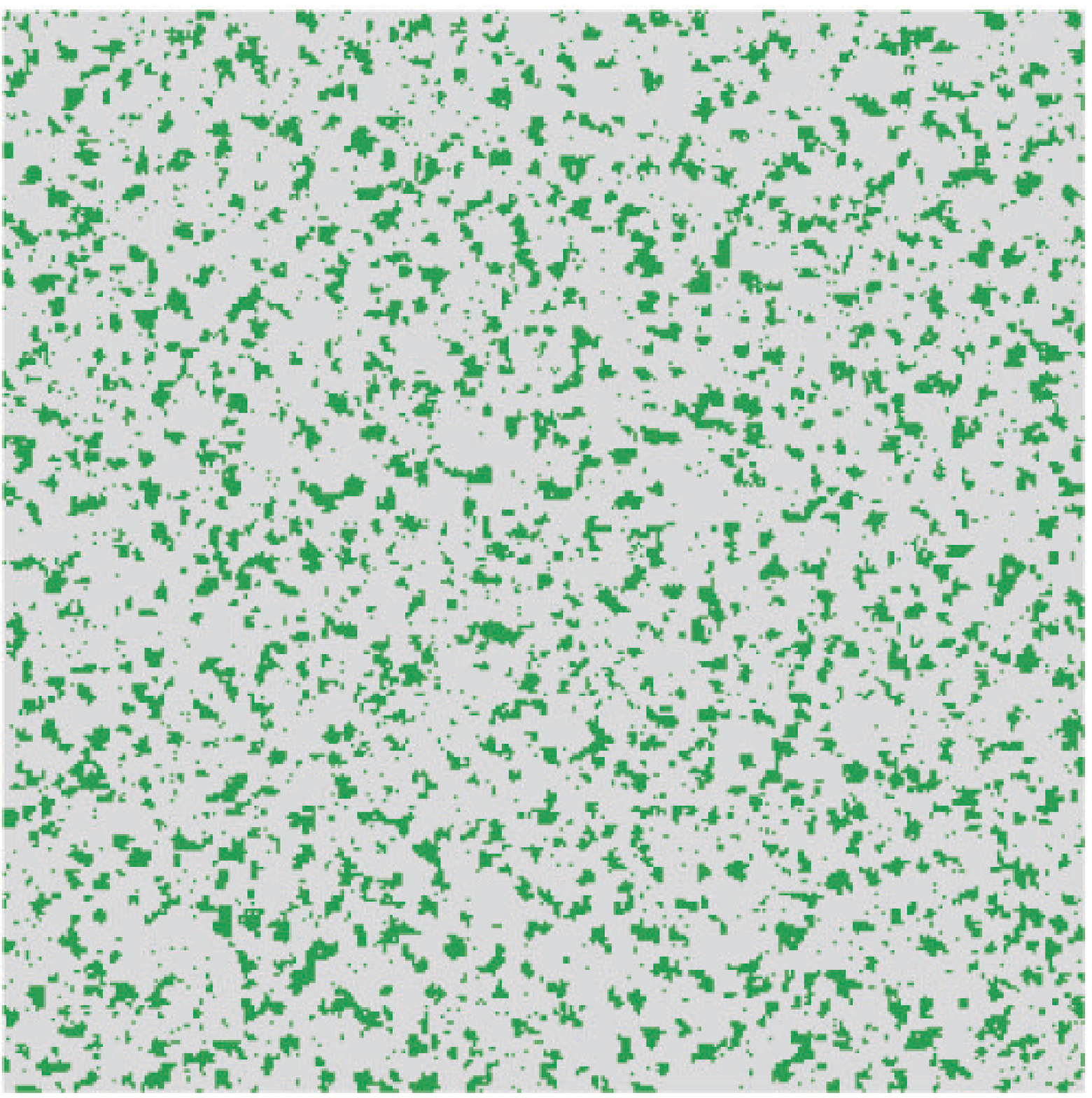}
\\
\includegraphics[width=6cm,clip=]{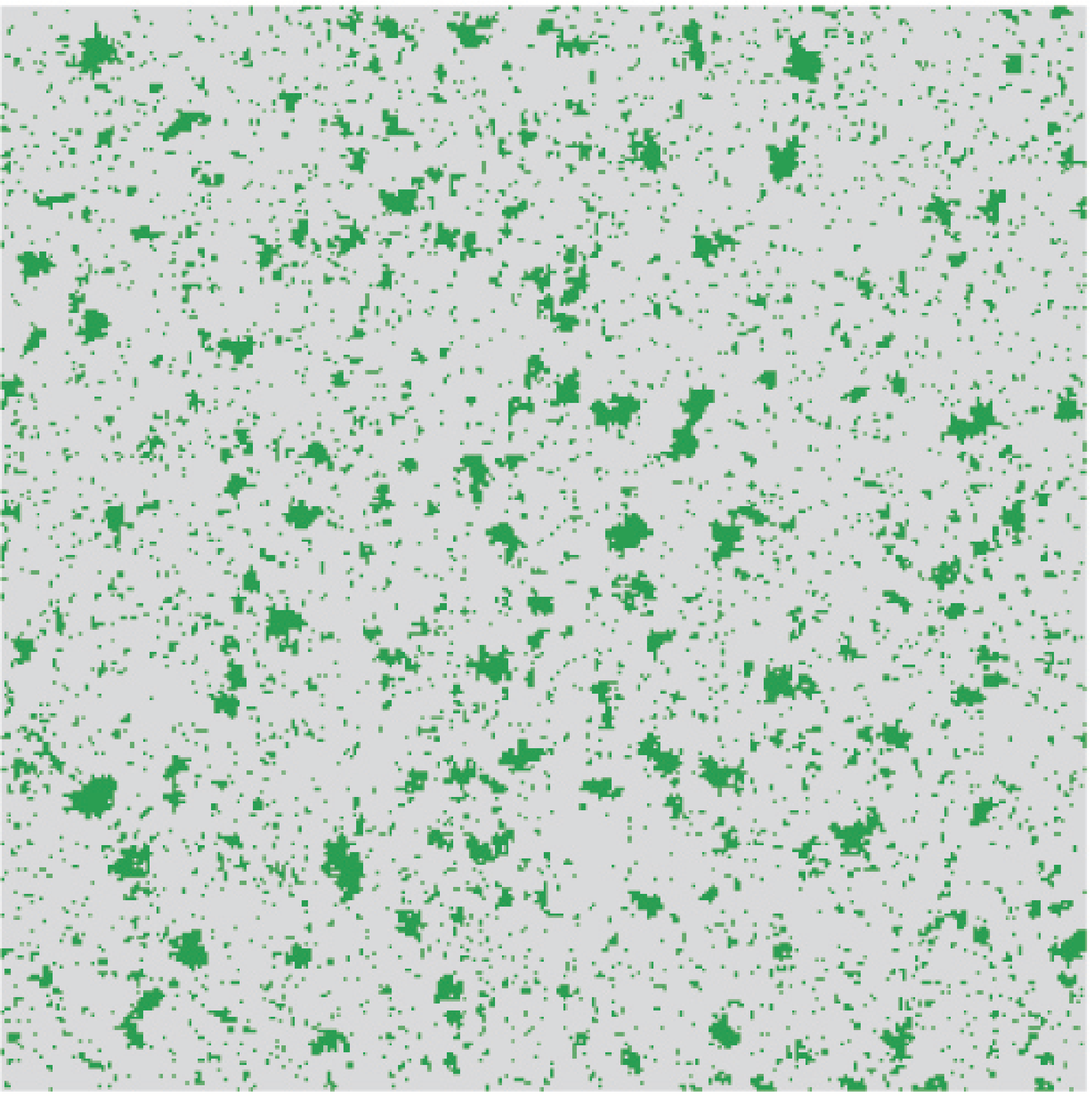}
\end{tabular}}
\caption{Clustering of $\Delta EGFR$ cells, as computed from the discrete model. Snapshots of the system are taken at time $T=1800$, which corresponds to the fifth day of experiments (see Fig.~\ref{snapshot}b). An upper panel corresponds to the case $\alpha_{high} = \alpha_{low}$, a lower panel incorporates an assumption that invasive cells switch their phenotype to proliferative (and increase their proliferation rate), when they form clusters, here $\alpha_{high} \gg \alpha_{low}$. Parameters: $q=0.9, \alpha_{low}=\alpha_{high}=0.00127$ for the upper panel and
$q=0.84, \alpha_{low}=0.0004, \alpha_{high}=0.0056$ for the lower panel, the diffusion time is $\tau = 4$ minutes.}
\label{simulations}
\end{figure}

Figure \ref{simulations} shows the results of simulations of the discrete model, which mimic the clustering behavior of
$\Delta EGFR$ cells, see  Fig. \ref{snapshot}b. The upper panel corresponds to the case $\alpha_{high} = \alpha_{low}$, the lower panel incorporates the assumption that invasive cells switch their phenotype to proliferative (and increase their proliferation rate), when they form clusters; here $\alpha_{high} \gg \alpha_{low}$. In the latter case, number of clusters is smaller, and the clusters are sufficiently large, which better agrees with experimental observations.

\begin{figure}[ht]
\centerline{\includegraphics[width=10cm,clip=]{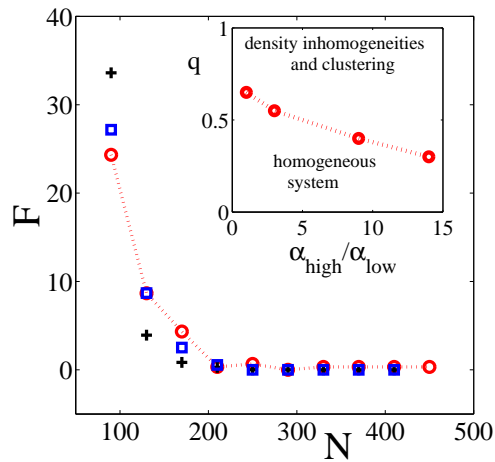}}
%\onefigure[width=7cm]{clustersize_final2.eps}
\caption{Cluster size distribution for $\Delta$EGFR from
experiment (circles, dotted line) and simulations (symbols). Pluses correspond to the case $\alpha_{high} = \alpha_{low}$, see also the upper panel in Fig. \ref{simulations}; squares correspond for the case we take into account the dynamic phenotypic switch between "invasive" phenotype and "proliferative" phenotype (see text) when cells form clusters, see also the lower panel in Fig. \ref{simulations}. Parameters are the same as in Fig. \ref{simulations}. An inset shows the phase diagram of parameters $(q, \alpha_{high})$ as computed from simulations of the discrete model for $\alpha_{low}=0.0017$ (see text). Above the dotted curve, the system shows density inhomogeneities, while below it the system remains homogeneous. Therefore, the possible parameters values for the wild type correspond to the lower part of the diagram.} \label{cluster_size}
\end{figure}

In order to make a quantitative comparison between the theoretical model and experiments, we consider the cluster size distribution, $F(N)$, for $\Delta$EGFR cells. It is hard to identify very small clusters in the experiment; thus we limit ourselves to the large $N$ tails of the distributions, see Fig. \ref{cluster_size}. We measured the size of sufficiently large clusters and averaged the results over three sets of experiments.  We estimated the number of particles in a cluster, assuming that the density within the cluster is about one-half the density of close packing. We performed a detailed comparison of the cluster size distribution obtained from numerical simulations with the experimental results both in case of "constant" proliferation ($\alpha_{high} = \alpha_{low}$) and in case where we distinguish between proliferative and invasive cells ($\alpha_{high} \gg \alpha_{low}$).

For ``constant" proliferation the agreement is rather poor (for any set of the unknown parameters $(q, \alpha)$). The number of small clusters in the simulations (Fig. \ref{cluster_size}, pluses) is larger than in the experiment (Fig. \ref{cluster_size}, circles, dotted line), while the number of large clusters is much smaller than in experiments. The same conclusion follows from the analysis of the corresponding snapshot of the system, Fig.~\ref{simulations}. The disagreement can be resolved by recalling that the proliferation rate is higher in the proliferative zone than in the invasive zone \cite{Giese}, $\alpha_{high} \gg \alpha_{low}$. Figure \ref{cluster_size} shows that the resulting cluster size distribution (Fig. \ref{cluster_size}, squares) is in a good agreement with the experiment on day $5$. This is also true on day $4$; it is quite difficult to make a comparison for earlier times, since the clusters are small. Therefore, our modeling suggests two successive mechanisms, leading to formation and growth of cell clusters. First, cells form small clusters due to the effects of cell-cell adhesion. Second, cells in these clusters switch their phenotype to proliferative and start dividing more rapidly.

As Fig. \ref{snapshot}a shows, wild-type cells do not form clusters. Nevertheless, it is possible to get some information about them by analyzing the total intensity of the green fluorescence in the image. This allows us to estimate the total cell density at early times, when the density growth is almost exponential. Therefore, we can derive the low-density proliferation rate, which turns out to be $\alpha_{low}=0.0017$. There are still two unknown parameters left for the wild type cells: the effective strength of cell-cell adhesion, $q$ and the division rate of proliferative cells, $\alpha_{high}$. The inset in Fig.~\ref{cluster_size} shows the phase diagram of parameters as computed from simulations of the discrete model. Each point in this phase diagram corresponds to the set of parameters $(q, \alpha_{high})$ for the fixed $\alpha_{low}$. Above the dotted curve, the system shows density inhomogeneities, while below the threshold cells remain distributed uniformly. Therefore, the possible parameters values for the wild type correspond to the lower part of the diagram.

\begin{figure} [ht]
%\vspace{-0.2cm}
\centerline{\includegraphics[width=9cm,clip=]{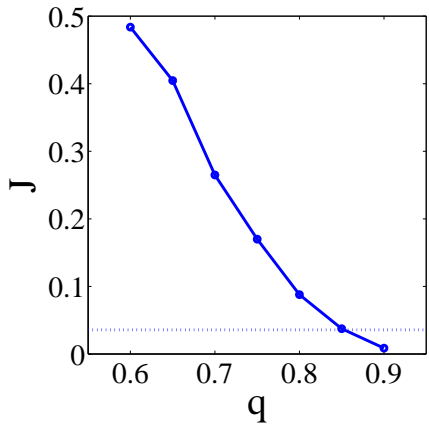}}
\caption{Shed rate $J$ from tumor spheroid (measured in $10^6$ $cells$ $cm^{-2}$ $day^{-1}$) as a function of
the adhesion parameter $q$ as computed in numerical simulations of a three-dimensional version of the discrete stochastic model. The dotted line corresponds to the experimentally observed shed rate for $U87 \Delta EGFR$ cells, which gives $q\simeq 0.85$.} \label{shedrate}
\end{figure}

In our modeling, the adhesion parameter $q$ determines the probability of detachment of a cell from a neighboring cell (or a cluster). How can this effective strength of cell-cell adhesion be experimentally measured? The value of $q$ for different cell types can be estimated by measuring the shed rate of invasive cells from the tumor surface. Since shedding is related to the process of breaking the cell-cell bond, by using the known shed rate and the tumor surface area, we arrive at values of the adhesion parameter $q$. We performed another experiment, in which $U87 \Delta EGFR$ tumor spheroid was placed in a collagen gel. Using confocal microscopy, we {\it counted directly} the number of cells detached from the spheroid surface per unit time per unit area. For the tumor with radius of $170\mu m$, we measured the shed rate of $0.0357 \times 10^6$ cells day$^{-1}$ cm$^{-2}$. The discrete model gives one-to-one correspondence between the shed rate $J$ and the adhesion parameter $q$, see Fig.~\ref{shedrate}. This gives $q\approx 0.85$ for $\Delta EGFR$ cells, which is in an excellent agreement with the values of $q$ used in our modeling. For the wild-type cells, we use estimates for the shed rate from \cite{Stein}, which give much smaller $q$.

In this work, we have studied the formation of clusters in a system of motile cells both experimentally and by simulations of a discrete stochastic model. We showed that there are two radically different regimes of behavior: cells either form clusters or remain homogeneously distributed over the system. We propose the following mechanism for cluster formation and growth. Initially small clusters are formed due to phase separation if cell-cell adhesion is larger than some threshold value. Then cells in these clusters switch their phenotype from ``invasive" to ``proliferative", which causes a rapid growth of clusters. Our experimental results are completely consistent with our theoretical predictions. The experiment shows that the process of phase separation, familiar in materials science, occurs in cell cultures.

The application of these ideas to \emph{in vivo} tumors is, necessarily, quite speculative, but not unrealistic. In fact, GBM and anaplastic astrocytomas are known to commonly form secondary foci of tumor formation within $2$ cm of the initially resected tumor. This would suggest that they behave as U87 $\Delta$EGFR cells with $q > q_c$.  However, another behavior seen in malignant gliomas is exemplified by the disease gliomatosis cerebri (GC). In GC, single malignant glioma cells infiltrate diffusely into the brain in a relatively homogeneous fashion. In this case, GC cells behave more like U87WT; their behavior is consistent with $q < q_c$. We should point out that secondary tumor formation is the obstacle to successful treatment of GBM patients after resection of the primary tumor. Cell-cell adhesion is known to be important in tumor growth and morphogenesis \cite{Takeichi}. We can guess that adhesion among the cloud of invasive cells (that are known to exist around GBM tumors) plays an important role in initial formation of clusters. We speculate that this may cause the dynamical switch of phenotype from "invasive" to "proliferative", which leads to development of recurrent tumors. Our model suggests that $\Delta$EGFR cells form clusters (if we return to the patterns in \cite{Stein,Khain0} we can suggest that the small clusters of $\Delta$EGFR cells are almost certainly due to enhanced adhesion in this cell type), which is consistent with the experimental observation that $\Delta$EGFR mutation is associated with higher malignancy \cite{EGFR}. Notice that these are a stable glioma cell line, they are basically a relatively uniform culture in which differentiation is not thought to be a major component. Investigating the overall effect of cell-cell adhesion on the morphology of a growing tumor \cite{Popawski} is an interesting avenue of future research.

Altered cell-cell adhesion is likely to be an important factor in glioma cell migration. One of the major mechanisms of cell-cell adhesion is the interaction of members of the cadherin family expressed on adjacent cells.  Cleavage of N-Cadherin has been shown to promote glioblastoma cell migration \cite{Kohutek}, and reorganization of the cadherins has been reported in glioma invasion \cite{Perego}. The importance of downregulation of cell-cell adhesion in glioblastoma invasion was emphasized by a study of the pro-adhesive gene OPCML, which is downregulated in glioma cell migration \cite{Reed}. Additional studies of biological cell-cell adhesion mechanisms in glioma will be necessary to understand the specific mechanisms pertinent to glioma cells.

Continuum modeling of cell-cell adhesion has attracted much recent attention \cite{Byrne,cont_adhesion}. A promising approach here is to proceed from the microscopic lattice models to a continuum description \cite{cont_adhesion}. A proper candidate for the continuum description of adhesive motile cells may be the Cahn-Hilliard equation, which is often used in condensed matter physics to describe the dynamics of phase separation below the critical temperature \cite{Godreche}. In our system, this corresponds to supercritical adhesion and zero proliferation, so an additional proliferation term should be included \cite{Khain2}. An interesting future research direction will be solving numerically the two-dimensional modified Cahn-Hilliard equation, deriving the cluster size distribution and comparing the results with the results of discrete modeling and in-vitro experiments.

Finally, we did not consider the effects of cell-matrix adhesion, however, it is no less important than cell-cell adhesion \cite{Hegedus1}. Cell-substrate adhesion is known to affect cell migration \cite{Zaman}. Therefore, choosing a specific migration time corresponds to the specific cell-substrate adhesion. In experiments, increase in cell-substrate adhesion may either increase or decrease the motility of cells, depending on functional ligand and receptor density \cite{Zaman}. It is widely believed that there is a competition between cell-cell and and cell-substrate adhesions, and that larger cell-substrate adhesion suppresses clustering \cite{Ryan}. Suppose that the density of ligands and receptors is such that increase in cell-substrate adhesion causes a decrease in cell motility. Then for supercritical adhesion, our model suggests that the larger the cell-substrate adhesion, the more time it would take to form clusters. This does not contradict the experimental findings of \cite{Ryan}, but suggests to wait for a longer time to see clustering. This may be an interesting avenue of future research.

We are grateful to Andrew M. Stein, Robert M. Ziff, Charles R. Doering, Thomas S. Deisboeck, and Eugene Surdutovich for useful discussions. Supported by NIH BRP grant R01 CA085139-01A2.


\begin{thebibliography}{0}
\bibitem{exp} The cells were stably transfected with plasmid pBos/H2B-EGFP which allows for mammalian expression of a histone H2B/EGFP fusion protein and cultured in standard Dulbeccos Modified Eagles Medium ($11995-065$, Invitrogen, Carlsbad, CA) with 10 percent addition of fetal bovine serum ($10438-026$, Invitrogen).
\end{thebibliography}
\end{document}